\documentclass[aps,prc,fleqn,floatfix,twocolumn,superscriptaddress,twoside]{revtex4}

\usepackage[dvipdfm]{graphicx}

\usepackage{amsmath,amssymb,amsfonts}

\usepackage{color}

\newcommand{\bea} {\begin{eqnarray}}
\newcommand{\eea} {\end{eqnarray}}

\newcommand{\mbx}{\mbox{}}

\begin{document}

\title{Parton Transport via Transverse and Longitudinal Scattering in Dense Media}

\author{Guang-You Qin}

\affiliation{Department of Physics and Astronomy, Wayne State University, Detroit, MI, 48201.}
\affiliation{Department of Physics, Duke University, Durham, NC, 27708.}

\author{Abhijit Majumder}

\affiliation{Department of Physics and Astronomy, Wayne State University, Detroit, MI, 48201.}

\date{\today}

\begin{abstract}

The effect of multiple scatterings on the propagation of hard partonic jets in a dense nuclear medium is studied in the framework of deep-inelastic scattering (DIS) off a large nucleus. Power counting arguments based on the Glauber improved Soft-Collinear-Effective-Theory are used to identify the class of
leading power corrections to the process of a single parton traversing the extended medium without emission. It turns out that the effect of longitudinal drag and
diffusion (often referred to as straggling) is as important as transverse scattering, when relying solely on power counting arguments.
With the inclusion of momentum exchanges in both transverse and longitudinal directions between the traversing hard parton and the constituents of the medium, we derive a differential equation for the time (or distance) evolution of the hard parton momentum distribution.
Keeping up to the second order in a momentum gradient expansion, this equation describes in-medium evolution of hard jets which experience longitudinal drag and diffusion plus the transverse broadening caused by multiple scatterings from the medium.

\end{abstract}
\maketitle

\section{Introduction}

The modification of hard jets in  hot (and or dense) extended media is a forefront topic in relativistic heavy-ion collisions as it may provide a sensitive measure of the properties of the produced highly excited matter~\cite{Wang:1991xy, Gyulassy:1993hr, Baier:1996kr, Zakharov:1996fv, Majumder:2010qh}.
The partonic showers initiated by these hard off-shell patrons are modified due to the scattering experienced by each of the partons in the
course of the propagation of the cascade through the medium.
Such a modification is confirmed by multiple experimental observations such as the significant depletion of high transverse momentum ($p_T$) hadron production compared to that in binary-scaled proton-proton collisions at the same energies \cite{Adcox:2001jp, Adler:2002xw, Aamodt:2010jd}.

The cause of this modification is a combination of drag, diffusion and, most importantly, stimulated emission experienced by the shower while propagating
through the medium \cite{Bjorken:1982tu, Braaten:1991we, Gyulassy:1999zd, Djordjevic:2003zk, Wiedemann:2000za, Salgado:2003gb, Arnold:2001ba, Moore:2004tg, Wang:2001ifa, Zhang:2003wk, Majumder:2009ge}.
Sophisticated quantitative jet quenching calculations have been performed for single inclusive hadron suppression \cite{Qin:2007rn, Bass:2008rv, Armesto:2009zi, Chen:2010te, Wicks:2005gt, Qin:2009gw},
as well as di-hadron correlations \cite{Zhang:2007ja, Renk:2008xq, Majumder:2004pt, Majumder:2004wh} and photon-hadron correlations \cite{Renk:2006qg, Zhang:2009rn, Qin:2009bk}.
Following a successful description of the data, a
significant amount of effort has now been placed on the quantitative extraction of jet transport parameters, such as $\hat{q} = d(\Delta p_T)^2/dt$ and $\hat{e} = dE/dt$, in hot and dense media.

The current manuscript represents an extension of the theory of parton energy loss,
based on the Glauber improved Soft-Collinear-Effective-Theory (GiSCET)
approach~\cite{Idilbi:2008vm, D'Eramo:2010ak}. This approach is closely linked with the
multiple scattering versions of the Higher Twist (msHT) scheme~\cite{Majumder:2007hx, Majumder:2009ge}
and our presentation will lie in-between these two approaches: we will not use the GiSCET Lagrangian; however, the power
counting rules of the different momentum components will be similar to those used in Ref.~\cite{Idilbi:2008vm}.

There is one exception to the above
statement which constitutes the \emph{raison d'\^{e}tre} of this paper. In all previous attempts to compute the effect of a
medium on the hard jet parton, the medium gluons are assumed to have vanishingly small light cone components, both in the
direction of the jet's momentum and in the direction opposite to it. In this work, as in Refs.~\cite{Majumder:2007hx,Majumder:2009ge},
the struck parton is assumed to possess a momentum $q = (q^{+}, q^{-}, q_{\perp}) \sim (\lambda^{2} Q, Q , \lambda Q)$, where
$Q$ is a hard scale and $\lambda$ is a small dimensionless number.  In the previous papers,
the gluons off which the jet scatters were assumed to have a momentum
$k \sim (\lambda^{2}, \lambda^{2}, \lambda) Q$. The transverse momentum components are required to have the same scale as those of the
jet. The $(+)$-components have to be $\sim \lambda^{2} Q$ to maintain the virtuality of the jet. However, there is no physical reason why
the $(-)$-component of $k$ has to be $\sim \lambda^{2} Q$. Indeed it may be as large as $\lambda Q$ without changing the virtuality of either
the hard parton or the Glauber gluon (there will be a difference in the effect on the medium).
In this paper, we explore the fate of a hard parton with propagates through a dense extended medium
interacting with such \emph{transverse and longitudinal Glauber} gluons.

In the current effort, the sole focus will lie on the effect of multiple scatterings; radiation and radiative energy loss for the
propagating hard parton are not included in this work.
{The longitudinal momentum exchange during multiple scatterings leads to the drag and diffusion of the propagating jet, which has often been understood as the collisional (or elastic) jet energy loss.
While the relative importance of the two mechanisms is still in dispute, collisional energy loss cannot simply be neglected. This is true, not only for the suppression of single inclusive light and heavy hadrons \cite{Qin:2007rn, Wicks:2005gt}, but also for jet shower evolution, energy loss distribution within and outside the jet cone, as well as energy and momentum deposition into the medium by the jet shower \cite{Qin:2009uh, Neufeld:2009ep, Qin:2010mn}. }
The longitudinal momentum exchange may affect induced radiation from the hard parton as well; this will
be presented in an upcoming publication.
The paper is organized as follows. In the next section, we present the three dimensional momentum distribution of the outgoing hard parton at leading twist. In Sec. III, we investigate the effect of multiple scatterings on this distribution. A class of the higher twist diagrams which are length enhanced are identified. Then the gradient expansion is invoked and the number of multiple scatterings is resummed. We finally derive a differential equation for the time evolution of parton momentum distribution as affected by multiple scatterings. Sec. IV contains our summary.

\section{Leading twist and parton distribution functions}

Consider the process of semi-inclusive DIS off a large nucleus where one hard parton with a momentum $l_q$ is produced,
\begin{eqnarray}
e(L_1) + A(P_A) \to e(L_2) + q(l_q)  + X
\end{eqnarray}
In the above expression, $L_1$ and $L_2$ represent the momenta of the incoming and outgoing leptons. The incoming nucleus of atomic number $A$ has a momentum $P_A=Ap$, with each nucleus carrying momentum $p$.
In the final state, all hadrons are detected and their momenta are summed to obtain the hard parton's momentum $l_q$.
$X$ denotes that such process is semi-inclusive.

Throughout this work, we utilize the light-cone component notations for four vectors ($p = [p^+, p^-, \vec{p}_\perp]$) with
\begin{eqnarray}
p^+ = (E+p_z)/\sqrt{2}, p^- = (E-p_z)/\sqrt{2}.
\end{eqnarray}
In the Briet frame of DIS, the incoming virtual photon $\gamma^*$ has a momentum $q$ and the nucleus has a momentum $P_A$,
\begin{eqnarray}
q = L_2 - L_1 = [-x_Bp^+, q^-, \vec{0}_\perp], P_A = A[p^+, 0, \vec{0}], \ \ \ \
\end{eqnarray}
where $x_B = Q^2 / (2p^+q^-)$ is the Bjorken variable in this frame.

The double differential cross section of the semi-inclusive process in which a jet with a momentum $l_q$ is produced may be expressed as
\begin{eqnarray}
\frac{E_{L_2} d\sigma}{d^3L_2 d^3 l_q} = \frac{\alpha_{e}}{2\pi s} \frac{1}{Q^4} L_{\mu\nu} \frac{dW^{\mu\nu}}{d^3l_q},
\end{eqnarray}
where $s=(p+L_1)^2$ is the total invariant mass of the lepton-nucleon collision system. The leptonic tensor is give by
\begin{eqnarray}
L_{\mu\nu} = \frac{1}{2} {\rm Tr}[\gamma \cdot L_1 \gamma_\mu \gamma \cdot L_2 \gamma_\nu].
\end{eqnarray}
The semi-inclusive hadronic tensor is defined as
\begin{eqnarray}
W^{\mu\nu} \!\!&&\!\! = \sum_{X} (2\pi)^4 \delta^4 (q + P_A - P_X) \nonumber\\ &&
\times \langle A| J^\mu(0)|X\rangle \langle X|J^\nu(0)|A\rangle.
\end{eqnarray}
In the above expression, $|A\rangle$ represents the initial state of an incoming nucleus with $A$ nucleons with momentum $p$ per nucleon, averaged over spins.
The state $|X\rangle$ represents the general final hadronic or partonic state, where the $\sum_X$ runs over all possible final states except the stuck hard jet.
$J^\mu = Q_q \psi_q \gamma^\mu \psi_q$ is the hadron electromagnetic current, with $Q_q$ the charge of a quark of flavor $q$ in units of the positron charge $e$.
Here our focus is on the final state interaction between the medium and the stuck quark and thus the discussion will be centered around the hadronic tensor; the leptonic tensor will not be discussed further.

\begin{figure}[thb]
\includegraphics[width=0.95\linewidth]{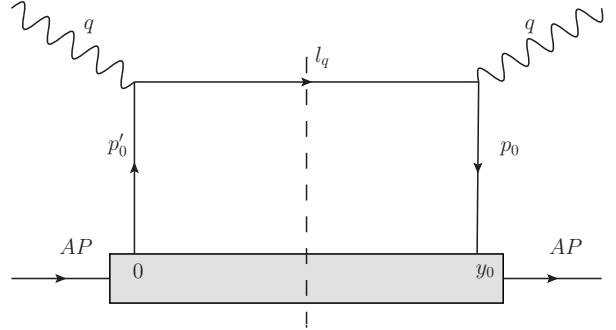}
 \caption{Leading twist contribution to the semi-inclusive DIS hadronic tensor $W^{\mu\nu}$.
} \label{leadingtwist}
\end{figure}

The leading twist contribution to semi-inclusive DIS is obtained by expanding the products of the currents at leading order in $\alpha_s$, as shown in Fig. \ref{leadingtwist}.
This represents the process where a hard quark, produced from one nucleon of the nucleus, exits the nucleus without further interaction.
The leading twist contribution to the semi-inclusive hadronic tensor may be expressed as
\begin{eqnarray}
\!\!&&\!\! W_0^{A \mu\nu}  = \sum_q Q_q^2  (-g_\perp^{\mu\nu}) A C_p^A  (2\pi) f_q(x_B).
\end{eqnarray}
In the above expression, The projection tensor $g_\perp^{\mu\nu}$ is defined as $g_\perp^{\mu\nu} = g^{\mu\nu} - g^{\mu-}g^{\nu+} - g^{\mu+} g^{\nu-}$.
The factor $C_p^A$ represents the probability to find a nucleon state with momentum $p$ inside a nucleus with $A$ nucleons.
The function $f_q(x)$ represents the parton distribution function of a quark with flavor $q$ in a nucleon with a fraction $x$ of the forward momentum $p^+$ of the nucleon,
\begin{eqnarray}
f_q(x) = \int \frac{dy_0^-}{2\pi} e^{-ixp^+y_0^-} \langle p| \bar{\psi}(y_0^-) \frac{\gamma^+}{2} \psi(0)|p \rangle,
\end{eqnarray}
where $\gamma^+$ and $\gamma^-$ are used to project out the leading helicity components,
\begin{eqnarray}
\gamma^+ = \frac{\gamma^0 + \gamma^3}{\sqrt{2}}, \gamma^- = \frac{\gamma^0 - \gamma^3}{\sqrt{2}}.
\end{eqnarray}

In the limit of very high forward momentum, the incoming parton carries vanishingly small transverse momentum ($p_{\perp} \sim \lambda Q$),
the differential hadronic tensor for the momentum distribution of the final quark may be approximated as,
\begin{eqnarray}
\frac{dW_0^{A\mu\nu}}{d^2l_{q \perp} dl_q^-}  = {W_0^{A\mu\nu}} \delta(l_q^- - q^-) \delta^2(\vec{l}_{q \perp}).
\end{eqnarray}
In the following section, we will investigate how the multiple scatterings from the medium change the momentum distribution of the final quark.

\section{Multiple scatterings and parton evolution in medium}

In this section, the effect of multiple scatterings on the 3-dimensional momentum distribution of the final quark is presented.
The higher twist contributions are obtained from those diagrams which include the expectation values of more partonic operators in the medium.
Such contributions are usually suppressed by powers of the hard scale $Q^2$,
but a sub-class of these contributions may be enhanced in extended media due to the longitudinal extent traveled by the struck quark.

The generic diagram being considered here is shown in Fig.~\ref{mntwist} which describes the process that a hard virtual photon strikes a hard quark in the nucleus with momentum $p_0'$ ($p_0$ in complex conjugate) at location $y_0'=0$ ($y_0$ in complex conjugate).
The stuck quark is then sent back through the nucleus and has momentum $q_1'$ ($q_1$ in the complex conjugate).
During its propagation through the nucleus, the hard parton scatters off the gluon field within the nuclear medium at locations $y_j'$ with $0<j<m$ ($y_i$ in the complex conjugate with $0<i<n$). In this effort, we are only considering the case where the hard quark propagates through the
medium without radiating. Radiation will be dealt with in a future effort.
Through each scattering, the hard parton picks up momenta $p_j'$ ($p_i$ in the complex conjugate).
Using the momentum conservation at each vertex, we may denote the various momenta in the picture as follows:
\begin{eqnarray}
&& q_{i+1} = q_i + p_i = q + \sum_{j=0}^i p_j = q + K_i, \nonumber\\
&& q'_{i+1} = q'_i + p'_i = q + \sum_{j=0}^i p'_j = q + K'_i,
\end{eqnarray}
where for convenience we have defined new momentum variables $K_i = \sum_{j=0}^{i} {p_j}$, and $K_i' = \sum_{j=0}^{i} {p_j'}$, which represent the total accumulated momentum exchanged between the hard quark and the nuclear medium.
For such a diagram, we may write down the hadronic tensor as,
\begin{widetext}
\begin{eqnarray}
W_{mn}^{A\mu\nu} \!\!&=&\!\! \sum_q Q_q^2 g^{n+m} \frac{1}{N_c} {\rm Tr}\left[\left(\prod_{i=1}^{n} T^{a_i}\right) \left(\prod_{j=m}^{1} T^{a'_j}\right) \right]
\int \frac{d^4l_q}{(2\pi)^4} (2\pi)\delta^{+}(l_q^2) \int d^4y_0 e^{iq\cdot y_0} \left(\prod_{i=1}^{n} \int d^4y_i\right) \left(\prod_{j=1}^{m} \int d^4y'_j \right)
\nonumber \\
&&  \left(\prod_{i=1}^{n} \int \frac{d^4q_i}{(2\pi)^4} e^{-iq_i\cdot (y_{i-1} - y_i)} \right) e^{-il\cdot (y_{n} - {y'}_{m})}  \left(\prod_{j=1}^{m}  \int \frac{d^4q'_j}{(2\pi)^4} e^{-
iq'_j\cdot (y'_j - {y'}_{j-1})}\right)
\nonumber \\
&&   \langle A | \bar{\psi}(y_0) \gamma^\mu \left(\prod_{i=1}^{n} \frac{\gamma\cdot q_i}{q_i^2 - i\epsilon} \gamma\cdot A^{a_i}(y_i)\right) \gamma\cdot l_q
\left(\prod_{j=m}^{1} \gamma\cdot A^{a_j'}(y_j') \frac{\gamma\cdot q'_j}{{q'_j}^2 + i \epsilon}\right) \gamma^\nu \psi(0) |A\rangle.
\end{eqnarray}
Here, $Q_{q}$ is the electromagnetic charge of the quark in units of an electron's charge, $N_{c}$ is the number of colors, and $T^{a_{i}} (T^{a'_{j}})$ represent Gell-Mann matrices.
To simplify the above expression, we may change the integral variables $q_{i+1} \to p_i$ and $q_{j+1}' \to p_j'$, and introduce the $n^{\rm th}$ exchanged momentum $p_n$ in the complex conjugate by inserting the following momentum conserving $\delta$-function,
\begin{eqnarray}
1 = \int \frac{d^4p_n}{(2\pi)^4} (2\pi)^4 \delta^4(l_q-q+K_n).
\end{eqnarray}
The $m^{\rm th}$ momentum $p_m'$ in the amplitude is determined by all other momenta, i.e., $p_m' = K_n - K_{m-1}'$.
The phase factor appearing in the above expression can be collected as $\left(\prod_{i=0}^n e^{-ip_i\cdot y_i} \right) \left(\prod_{j=0}^m e^{ip'_j\cdot y'_j} \right)= \left(\prod_{i = 0}^{n} e^{-ip_i \cdot (y_i-y'_{m})} \right) \left(\prod_{j=0}^{m-1} e^{ip'_j \cdot (y'_j-y'_{m})}  \right)$.

\begin{figure}[thb]
\includegraphics[width=0.95\linewidth]{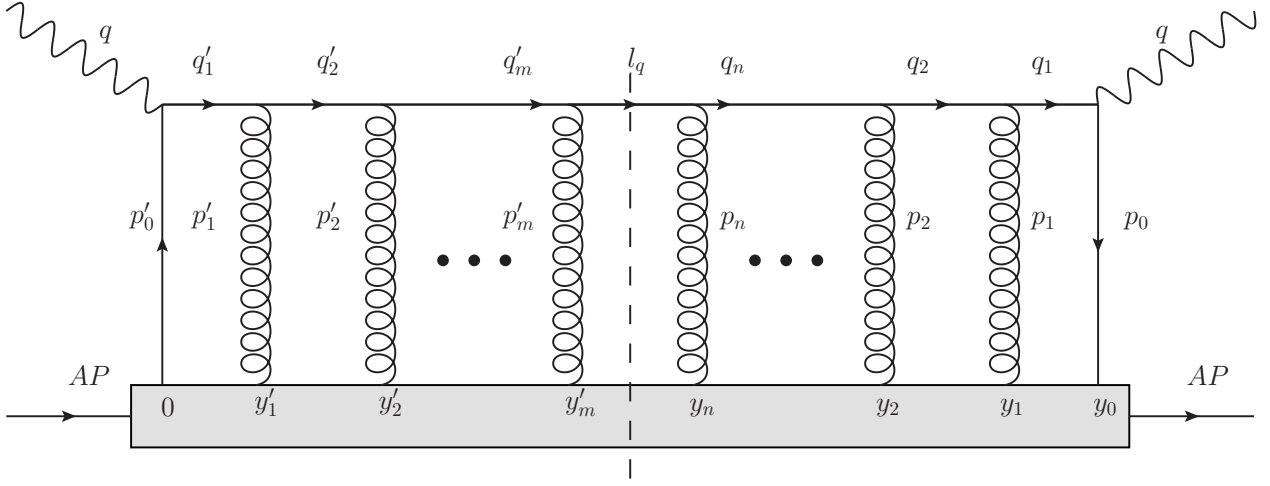}
 \caption{An order of $n+m$ contribution to hadronic tensor $W^{\mu\nu}$ with $n$ gluon insertions in the complex conjugate and $m$ gluon insertions in the amplitude.
} \label{mntwist}
\end{figure}

Before progressing further, we outline the power counting relations which underly the various approximations that will be carried out
in the remainder of this paper.
The momenta exchanged with the medium has transverse components that scale as $k_{\perp} \sim \lambda Q$ where $Q$ represents
 the hard scale in the problem. Different from prior calculations we will also insist that $k^{-} \sim \lambda Q$. Given the large Lorentz
 boost, the $(+)$-component of $k$ may be quite large, however, due to the requirement that the propagating parton not go off-shell by more than
 $\lambda Q$, we will obtain that $k^{+} \sim \lambda^{2} Q$.
We now consider the propagators after each scattering.
For the quark momentum after $i^{\rm th}$ scattering, we have
\begin{eqnarray}
q_{i+1}^2  = (q+K_i)^2 = 2p^+ q^- (1 + K_i^-/q^-) \left[ -x_B + \bar{x}_i - { \bar{x}_{D} \mbx}_i \right],
\end{eqnarray}
where we have defined a few momentum fraction variables,
\begin{eqnarray}
\bar{x}_i = \sum_{j=0}^i x_i = \sum_{j=0}^i \frac{p_j^+}{p^+} = \frac{K_i^+}{p^+}, \ \ \ {\bar{x}_{D} \mbx}_i = \sum_{j=0}^i {x_{D} \mbx}_i = \frac{K_{i,\perp}^2}{2p^+q^-(1 + K_i^-/q^-)}.
\end{eqnarray}
We may collect the contributions from all the internal quark line denominators together with the on-shell delta function for the final quark line $l_q$ as,
\begin{eqnarray}
D_q \!\!&&\!\! = \frac{C_q }{(2p^+q^-)^{n+m+1}}
\left( \prod_{i=0}^{n-1} \frac{1}{-x_B+\bar{x}_i-{\bar{x}_{D} \mbx}_i} \right)
\left( \prod_{j=0}^{m-1} \frac{1}{-x_B + {\bar{x}' \mbx}_i - { { \bar{x}}'_{D} \mbx}_i} \right) (2\pi) \delta(-x_B+ \bar{x}_n - { \bar{x}_{D} \mbx}_n ),
\end{eqnarray}
where the factor $C_q$ stands for,
\begin{eqnarray}
C_q \!\!&&\!\!= \left( \prod_{i=0}^{n} \frac{1}{1+K_i^-/q^-} \right)
\left( \prod_{j=0}^{m-1} \frac{1}{1+{K'_j}^-/q^-} \right)  \label{C-q}.
\end{eqnarray}
In so doing, we have retained the first correction in $\lambda$ stemming from the
longitudinal loss experienced from exchanges with $k^{-} \sim \lambda Q$.
As argued in Ref.~\cite{Majumder:2007ne}, in the high energy ($Q \rightarrow \infty$) and collinear $(\lambda \rightarrow 0)$ limit,
we may drop the transverse parts of the $\gamma$ matrices which connect the quark current to the virtual photon, which implies,
\begin{eqnarray}
\langle \bar{\psi}(y) \hat{O} \psi(0) \rangle \approx \frac{\gamma^-}{2}  \langle \bar{\psi}(y) \frac{\gamma^+}{2} \hat{O}  \psi(0) \rangle.
\end{eqnarray}
Also, noting that $A^{+} \sim \lambda^{2} Q$ and $A_{\perp} \sim \lambda^{3} Q$, in $A^{-} = 0$ gauge~\cite{Majumder:2009ge}, we approximate,
\begin{eqnarray}
\gamma \cdot A(y) \approx \gamma^- A^+(y). \label{A+}
\end{eqnarray}
One should point out, once again, that in this manuscript we will demonstrate that the effect of longitudinal drag is the same order as that of
transverse scattering; however, in the derivation, we will retain corrections up to order $\lambda$.
While the order $\lambda$ corrections to the propagators are retained in Eq.~\eqref{C-q}, we neglect
contributions to the vertices from $A_{\perp}$ in Eq.~\eqref{A+}. This is due to the fact that the first non-zero
contribution containing an $A_{\perp}$ term is of the from $\gamma_{\perp} \cdot {q_{i}}_{\perp} \gamma_{\perp} \cdot A_{\perp}$ which is
of order $\lambda^{2}$.

With these simplifications, we obtain
\begin{eqnarray}
\!\!\! && \langle A | \bar{\psi}(y_0) \gamma^\mu \left(\prod_{i=1}^{n} \gamma\cdot q_i \gamma\cdot A^{a_i}(y_i) \right) \gamma \cdot l_q \left(\prod_{j=m}^{1}
\gamma \cdot  A^{a'_j} (y'_j) \gamma\cdot q'_j\right) \gamma^\nu \psi(0) |A\rangle \nonumber\\
\!\!\! &&= \langle A | \bar{\psi}(y_0) \frac{\gamma^+}{2} \left(\prod_{i=1}^{n} A^{+ a_i}(y_i)\right) \left(\prod_{j=m}^{1} A^{+ a'_j}(y'_j)\right) \psi(0) |A\rangle
{\rm Tr} \left[\frac{\gamma^-}{2} \gamma^\mu \left(\prod_{i=1}^{n} \gamma\cdot q_i \gamma^{-}\right) \gamma
\cdot l_q \left(\prod_{j=m}^{1} \gamma^{-} \gamma\cdot q'_j\right) \gamma^\nu \right]. \ \ \ \ \ \
\end{eqnarray}
We now simplify the trace over spinor indices. Using $q_{i+1} = q + K_i = q + \sum_{j=0}^i p_i$, the trace reads,
\begin{eqnarray}
T = {\rm Tr}\left[\frac{\gamma^-}{2} \gamma^\mu \left(\prod_{i=1}^{n}
\gamma \cdot (q+ K_{i-1}) \gamma^{-} \right) \gamma \cdot (q+K_n) \left(\prod_{j=m}^{1}
\gamma^{-} \gamma \cdot (q+K_{j-1}')\right) \gamma^\nu \right],
\end{eqnarray}
where $\gamma \cdot (q+K_i)  = \gamma^+(q^- + K_i^-)  + \gamma^-(-x_B p^+ + K_i^+) - \vec{\gamma}_\perp \cdot \vec{K}_{i, \perp}$.
The above trace can be easily carried out by noting that $\gamma^-\gamma^-=0$, as a result, all terms in $\gamma \cdot (q+K_i)$ containing a
$\gamma^-$ vanish. Terms containing $\gamma_{\perp} \cdot K_{\perp}$ can only be included along with a $\gamma_{\perp} A_{\perp}$ and thus
contribute $\lambda^{2}$ corrections to the cross section. As a result, we obtain,
\begin{eqnarray}
T = (- g_\perp^{\mu\nu})  \frac{{(2q^-)}^{n+m+1}}{C_q} .
\end{eqnarray}
Substituting the above simplifications into the hadronic tensor, one obtains,
\begin{eqnarray}
W_{mn}^{A\mu\nu} \!\!&=&\!\! \sum_q Q_q^2 g^{n+m} \frac{1}{N_c} {\rm Tr} \left[ \left( \prod_{i=1}^{n} T^{a_i} \right) \left( \prod_{j=m}^{1} T^{a'_j} \right) \right]
\int \frac{d^3l_q}{(2\pi)^3}  (2\pi)^3 \delta^3(\vec{l}_q-\vec{q}-\vec{K}_n) \nonumber \\
&& \left(\prod_{i=0}^{n} \int dy_i^- \int d^3y_i \right) \left(\prod_{j=1}^{m} \int d{y'}_j^- \int d^3y'_j\right)
\left(\prod_{i=0}^{n} \int \frac{dx_i}{2\pi} \int \frac{d^3p_i}{(2\pi)^3}\right)
\left(\prod_{j=0}^{m-1}  \int \frac{{dx'}_j}{2\pi} \int \frac{d^3p'_j}{(2\pi)^3} \right) \nonumber \\
&& \left(\prod_{i = 0}^{n} e^{-ix_i p^+ (y_i^--{y'}_{m}^-)} e^{-i \vec{p}_i \cdot
(\vec{y}_i - \vec{y}'_{m})} \right) \left(\prod_{j=0}^{m-1} e^{i{x'}_j^+ p^+ ({y'}_j^--{y'}_{m}^-)}
e^{i \vec{p}'_j \cdot (\vec{y}'_j - \vec{y}'_{m})} \right) \nonumber \\
&&   (2\pi) \delta(- x_B + \bar{x}_n - {\bar{x}_{D} \mbx}_n)\left(\prod_{i=0}^{n-1} \frac{1}{-x_B + \bar{x}_i - {\bar{x}_{D} \mbx}_i - i\epsilon}\right)
\left(\prod_{j=0}^{m-1} \frac{1}{-x_B + {\bar{x'}}_j - {\bar{x_{D}'}}_j + i\epsilon}\right)  \nonumber\\
&& (-g_\perp^{\mu\nu}) \langle A | \bar{\psi}(y_0) \frac{\gamma^+}{2}
\left(\prod_{i=1}^{n} A^{+ a_i}(y_i)\right) \left(\prod_{j=m}^{1} A^{+ a'_j}(y'_j)\right) \psi(0) |A\rangle,
\end{eqnarray}
where we have performed the integration over ${l^+}$ using the delta function $(2\pi) \delta(l^+ - q^+ - K_n^+)$,
and have changed integration variables $p_i^+ \to x_i = p_i^+/p^+$, $p_j^{+'} \to x_j' = p_j^{+'}/p^+$.
We have introduced a vector notation for momentum and space coordinates for convenience:
$\vec{p} = (p^-, \vec{p}_\perp)$ and $\vec{y} = (y^+, \vec{y}_\perp)$,
and their dot product represents $\vec{p}\cdot \vec{y} = p^-y^+ - \vec{p}_\perp \cdot \vec{y}_\perp$.

Now we perform the integration over the momentum fractions $x_i$ and $x_j'$.
The integration over $x_n$ is performed using the last delta function which forces the cut line to be on shell and yields,
\begin{eqnarray}
x_n = -\bar{x}_{n-1} + x_B + {\bar{x}_{D} \mbx}_n = -\sum_{i=1}^{n-1} x_i + x_B + {\bar{x}_{D} \mbx}_n.
\end{eqnarray}
The (+)-component of the phase factor now reads,
\begin{eqnarray}
\Gamma^+ = e^{-i(x_B + {\bar{x}_{D} \mbx}_n)p^+y_n^-} \left(\prod_{i = 0}^{n-1} e^{-ix_i p^+ (y_i^--y_n^-)}\right)
e^{i(x_B + \bar{x_{D}' \mbx}_{m})p^+{y'}_{m}^-} \left(\prod_{j=0}^{m-1} e^{i{x'}_j p^+ ({y'}_j^--{y'}_{m}^-)} \right) = \Gamma_n^+ \Gamma_m^+ .
\end{eqnarray}
In the above equation, $\Gamma_n^+$ and $\Gamma_m^+$ represent the two phase factors associated with the $x_i$ integration and
the $x'_j$ integration, respectively.
The remaining integrations over the momentum fractions $x_i$ and $x'_j$ may now be performed using contour integration.
One starts from the propagators adjacent to the cut and proceeds to the initial hard scattering vertex.
In the following, the integrations over the fractions in the complex conjugate ($x_i$'s) will be described in detail;
the integrations over the fractions $x'_j$'s in the amplitude are completely analogous.

The first integration is the one over the momentum fraction $x_{n-1}$.
Isolating the piece related to $x_{n-1}$ integral, we can perform the integral by closing the contour of $x_{n-1}$ with a counter-clockwise semi-circle in the upper half of  $x_{n-1}$ complex plane and obtain
\begin{eqnarray}
\int \frac{dx_{n-1}}{2\pi} \frac{e^{-ix_{n-1} p^+ (y_{n-1}^{-} -y_n^-)}}{-x_B + x_{n-1}
+ \bar{x}_{n-2} - {\bar{x}_{D} \mbx}_{n-1} - i\epsilon} = i \theta(y_n^--y_{n-1}^-) e^{-i (-\bar{x}_{n-2} + x_B + {\bar{x}_{D} \mbx}_{n-1})p^+ (y_{n-1}^- - y_n^-)}.
\end{eqnarray}
Here the $\theta$-function represents the physical effect that the quark line is propagating from $y_{n-1}^-$ to $y_n^-$.
Correspondingly, the phase factor combined with the above from the contour integration becomes,
\begin{eqnarray}
\Gamma_n^+ \to e^{-i {x_{D} \mbx}_n p^+y_n} e^{-i(x_B + {\bar{x}_{D} \mbx}_{n-1}) p^+ y_{n-1}^-} \left(\prod_{i = 0}^{n-2} e^{-ix_i p^+ (y_i^--y_{n-1}^-)}\right).
\end{eqnarray}
The phase factor is found to have a structure similar to the case without longitudinal drag~\cite{Majumder:2007hx}.
Thus, the remaining integrals over the longitudinal momentum fractions $x_i$'s can be done in a similar way. Finally, we obtain
\begin{eqnarray}
\Gamma_n^+ \to e^{-i x_B p^+ y_0^-}  \left(\prod_{i=1}^n e^{-i {x_{D}}_i p^+ y_i^-} \right)
 i^n \left(\prod_{i=1}^n \theta(y_i^--y_{i-1}^-) \right).
\end{eqnarray}
The integrations over the momentum factions $x'_j$'s in the amplitude are completely analogous, except that a factor of $(-i)$ instead of $i$ is associated with the $\theta$ function. Such difference originates from the contour integral of $x_{n-1}$ with a clockwise semi-circle in the lower half of the complex plane.

After having performed the integrations over the internal quark lines, we may combine various parts and obtain the hadronic tensor as follows,
\begin{eqnarray}
W_{mn}^{A\mu\nu} \!\!&=&\!\! \sum_q Q_q^2 g^{n+m} \frac{1}{N_c} {\rm Tr}\left[\left(\prod_{i=1}^{n} T^{a_i}\right) \left(\prod_{j=m}^{1} T^{a'_j}\right) \right]
\int \frac{d^3l_q}{(2\pi)^3}  (2\pi)^3 \delta^3(\vec{l}_q-\vec{q}-\vec{K}_n) \nonumber \\
&& \left(\prod_{i=0}^{n} \int dy_i^- \int d^3y_i \right) \left(\prod_{j=1}^{m} \int d{y'}_j^- \int d^3y'_j\right) \left(\prod_{i=0}^{n} \int \frac{d^3p_i}{(2\pi)^3}\right)
\left(\prod_{j=0}^{m-1}  \int \frac{d^3p'_j}{(2\pi)^4} \right)
\left(\prod_{i = 0}^{n} e^{-i \vec{p}_i \cdot \vec{y}_i } \right) \left(\prod_{j=1}^{m}  e^{i \vec{p}'_j \cdot \vec{y}'_j} \right)  \nonumber\\
&&  e^{-i x_B  p^+ y_0^-}  \left(\prod_{i=1}^n e^{-i {x_{D}}_i p^+ y_i^-} \right) \left(\prod_{j=1}^{m} e^{i {x_{D}'}_j p^+ {y'}_j^-} \right) i^n (-i)^{m} \left(\prod_{i=1}^n \theta(y_i^--y_{i-1}^-) \right) \left(\prod_{j=1}^{m} \theta({y'}_j^- - {y'}_{j-1}^-) \right) \nonumber\\
&& (-g_\perp^{\mu\nu})
\langle A | \bar{\psi}(y_0) \frac{\gamma^+}{2} \left(\prod_{i=1}^{n} A^{+ a_i}(y_i)\right) \left(\prod_{j=m}^{1} A^{+ a'_j}(y'_j)\right) \psi(0) |A\rangle.
\end{eqnarray}
The expression derived above is quite general in the sense that we have not made any assumption regarding the nature of the nuclear states.
In the following, we will make specific assumptions on the nuclear states to simplify the above expression.
Here we only consider the case of $n=m$, i.e., the same number of scatterings in both the amplitude and the complex conjugate.
The case where $n\ne m$, either constitute higher twist nucleon matrix elements or unitarity corrections to terms where the produced quark experiences scatterings $\min(n,m)$ times \cite{Majumder:2007hx}.

For the case of $2n$ ($n=m$) gluon insertions, we may first simplify the matrix elements of the gluon vector potentials in the nuclear state.
In this work, the nucleus is approximated as a weakly interacting homogenous gas of nucleons.
Such approximation is sensible at very high energy, where nucleons appear to travel in straight lines and are almost independent of each other over the time interval of the interaction of the hard probe due to time dilation.
As a result, we may decompose the expectation of field operators in the nuclear states into the expectation in the nucleon states.
As a nucleon is a color singlet, any combination of quark or gluon field strength insertions must be restricted to a color singlet.
Therefore, the first non-zero (and the largest) contribution comes from the terms where $2n$ gluons are divided into singlet pairs in separate nucleon states where one gluon of the pair is in the amplitude and one is in the complex conjugate.

Using the time ordered products of $\theta$ functions above, the expectation of the field operators in the nuclear state may be decomposed as,
\begin{eqnarray}
\langle A | \bar{\psi}(y_0) \frac{\gamma^+}{2} \left(\prod_{i=1}^{n} A^{+ a_i}(y_i)\right) \left(\prod_{j=n}^{1} A^{+ a'_j}(y'_j)\right) \psi(0) |A\rangle = C_{p_0, p_1 \cdots p_n} \langle p | \bar{\psi}(y_0) \frac{\gamma^+}{2} \psi(0) |p\rangle \left(\prod_{i=1}^{n} \langle p| A^{+ a_i}(y_i) A^{+ a'_i}(y'_i)|p\rangle \right). \nonumber\\
\end{eqnarray}
In the above decomposition, we keep only the largest contribution arising from the terms where the expectation of each pair of partonic operators is evaluated in separate nucleon states. The $y_i$ integrations are carried out over the nuclear volume under the constraints
imposed by the string of $\theta$-functions.
The factor $C_{p_0, p_1 \cdots p_n}$ represents the probability to find $n+1$ nucleons in the vicinity of the positions $y_0, y_1 \cdots y_n$. In the case of non-interacting nucleons in a nucleus with a uniform density, this factor may be approximated as
\begin{eqnarray}
C_{p_0, p_1 \cdots p_n} = A C_p^A \left(\frac{\rho}{2p^+}\right)^n ,
\end{eqnarray}
where $\rho$ is the nucleon density inside the nucleus, and the factor $1/(2p^+)$ originates in the normalization of nucleon state.
Note that in the case of a nucleus which does not have a uniform density the coefficient will possess a spatial dependence.
One may average over the colors of the gluon fields,
\begin{eqnarray}
\langle A^a(y) A^b(0)\rangle = \frac{\delta_{ab}}{N_c^2 - 1} \langle A(y) A(0)\rangle.
\end{eqnarray}
The average over the colors of quark field has already been carried out to give the factor $1/N_c$.
Now the overall trace over the color factors reduces to
\begin{eqnarray}
\frac{1}{N_c} \frac{1}{(N_c^2 -1)^n} {\rm Tr}\left[\left(\prod_{i=1}^{n} T^{a_i}\right) \left(\prod_{j=n}^{1} T^{a_i}\right) \right]  = \left(\frac{C_F}{N_c^2-1}\right)^n.
\end{eqnarray}
We may now obtain the hadronic tensor as
\begin{eqnarray}
W_{nn}^{A\mu\nu} \!\!&=&\!\! \sum_q Q_q^2 (-g_\perp^{\mu\nu})
 A C_p^A \left(\frac{\rho}{2p^+}\right)^n g^{2n} \left(\frac{C_F}{N_c^2 - 1} \right)^n
\int \frac{d^3l_q}{(2\pi)^3}  (2\pi)^3 \delta^3(\vec{l}_q-\vec{q}-\vec{K}_n) \nonumber \\
&& \int dy_0^- \int d^3y_0 \int \frac{d^3p_0}{(2\pi)^3} e^{-i \vec{p}_0 \cdot \vec{y}_0} e^{-i x_B p^+ y_0^-}  \langle p | \bar{\psi}(y_0) \frac{\gamma^+}{2} \psi(0) |p\rangle \nonumber \\
&&  \left(\prod_{i=1}^{n} \int dy_i^- \int d{y'}_i^- \theta(y_i^--y_{i-1}^-) \theta({y'}_i^- - {y'}_{i-1}^-) \right) \nonumber \\
&& \left(\prod_{i=1}^{n} \int d^3y_i \int d^3y'_i \int \frac{d^3p_i}{(2\pi)^3} \int \frac{d^3p'_i}{(2\pi)^3} e^{-i \vec{p}_i \cdot \vec{y}_i }  e^{i \vec{p}'_i \cdot \vec{y}'_i} e^{-i {x_{D}}_i p^+ y_i^-}  e^{i {x_{D}'}_i p^+ {y'}_i^-} \langle p| A^{+}(y_i) A^{+}(y'_i)|p\rangle \right),
\end{eqnarray}
where we have changed variables from $dp'_0$ to $dp'_n$.

The above expression may be further simplified by changing variables $(y_i, y'_i) \to (Y_i, \delta y_i)$,
\begin{eqnarray}
Y_i = (y_i + y'_i)/2, \ \ \ \delta y_i = y_i - y'_i .
\end{eqnarray}
For a large enough nucleus one may insist on translational invariance for the expectation values of gluon operators, i.e.,
\begin{eqnarray}
\langle p| A^{+}(y_i) A^{+}(y'_i)|p\rangle \simeq \langle p|A^+(y_i-y'_i) A^+(0) |p\rangle \simeq \langle p|A^+(\delta y_i) A^+(0) |p\rangle .
\end{eqnarray}
Now only the phase factor depends on the average values $\vec{Y}_i$, thus we may perform the integration over the phase factor and obtain
\begin{eqnarray}
\int d^3y_i \int d^3y'_i  e^{-i \vec{p}_i \cdot \vec{y}_i }  e^{i \vec{p}'_i \cdot \vec{y}'_i}  = \int d^3Y_i \int d^3 \delta y_i e^{-i (\vec{p}_i-\vec{p}'_i) \cdot \vec{Y}_i } e^{-i (\vec{p} + \vec{p}'_i) \cdot \vec{\delta y}_i/2}  = (2\pi)^3 \delta^3(\vec{p}_i-\vec{p}'_i) \int d^3 \delta y_i e^{-i \vec{p} \cdot \vec{\delta y}_i}.
\end{eqnarray}
The above delta function will set the momentum fractions ${x_{D}}_i = {x_{D}'}_i$.
For the product of $\theta$-functions, since both $\delta y_i^-$ and $\delta y_{i-1}^-$ are within the nucleon size (small compared to the nucleus size $Y^-$), we may simplify it as,
\begin{eqnarray}
\theta(y_i^--y_{i-1}^-) \theta({y'}_i^- - {y'}_{i-1}^-) = \theta(Y_i^- - Y_{i-1}^-).
\end{eqnarray}
The time-ordered product of $\theta$-functions gives,
\begin{eqnarray}
\left(\prod_{i=1}^{n} \int_0^{L^-} dY_i^- \theta(Y_i^- - Y_{i-1}^-)\right)  = \frac{1}{n!} \left(\prod_{i=1}^{n} \int_0^{L^-} dY_i^- \right),
\end{eqnarray}
where $L^-$ is the extent of the nucleus size.
Combining the above simplifications and using the delta function to perform the integration over $p'_i$, we obtain the differential hadronic tensor as,
\begin{eqnarray}
\frac{d W_{nn}^{A\mu\nu}}{{d^3l_q} } \!\!&=&\!\! W_0^{A\mu\nu}  \frac{1}{n!} \prod_{i=1}^{n} \left( \int_0^{L^-} dY_i^- \int d\delta y_i^- \int d^3 \delta y_i \int \frac{d^3p_i}{(2\pi)^3} \frac{\rho}{2p^+} g^{2} \frac{C_F}{N_c^2 - 1}
e^{-i \vec{p}_i \cdot \delta \vec{y}_i} \langle p| A^{+}(\delta y_i) A^{+}(0) |p\rangle \right) \nonumber \\
&&  \left(\prod_{i=1}^{n}  e^{-i {x_{D}}_i p^+ \delta y_i^-}  \right) \delta^3 \left( \vec{l}_q - \vec{q} - \sum_{i=1}^n \vec{p}_i \right).
\end{eqnarray}

With all the simplifications carried out above, it is now possible to carry out the resummation over multiple scatterings.
As in the case without longitudinal exchange, we will not resum the entire hadronic tensor as outlined above. Instead we introduce further
simplifications arising from the collinear approximation, which restricts the exchanged momenta to be small compared to the energy of the
hard parton.
One Taylor expands the hard part (the last line in the above expression) in terms of the exchanged momenta $\vec{p}_i$ around $\vec{p}_i = 0$,
\begin{eqnarray}
H = \left(\prod_{i=1}^{n} \left[1 + p_i^\alpha \frac{\partial}{\partial p_i^\alpha} + \frac{1}{2} p_i^\alpha p_i^\beta \frac{\partial}{\partial p_i^\alpha} \frac{\partial}{\partial p_i^\beta} + \cdots \right] \right) H|_{\vec{p}_1 \cdots \vec{p}_n = 0}.
\end{eqnarray}
In the above equation, the values of $\alpha, \beta$ take the values of ``$-$'' and ``$\perp$''.
Here we only keep the terms in the expansion up to the second order derivatives,
with the assumption of small momentum exchange in each of the multiple scatterings.
The extension to include higher derivative terms should be straightforward,
which corresponds to including higher moments of the momentum distribution of the exchanged gluon.
In the above expansion, these terms involving the zeroth order term (without derivative) represent the case where the exchanged gluon carries zero momentum. Such terms corresponds to gauge corrections to diagrams with lower number of scatterings  and will not be considered further. These terms
should be understood to be included in the gauge invariant definition of the transport coefficients.

Using integration by parts, the factors of exchanged momentum may be converted into the appropriate derivatives over
relative position,
\begin{eqnarray}
\!\!&&\!\! e^{-i\vec{p}_i \cdot \delta\vec{y}_i} \langle p| A^+(\delta \vec{y}_i) A^+(0)|p \rangle p_i^\alpha \frac{\partial}{\partial p_i^\alpha} =   e^{-i\vec{p}_i \cdot \delta\vec{y}_i} (-i) \langle p| \partial^\alpha A^+(\delta \vec{y}_i) A^+(0)|p \rangle \frac{\partial}{\partial p_i^\alpha} , \nonumber\\
\!\!&&\!\! e^{-i\vec{p}_i \cdot \delta\vec{y}_i} \langle p| A^+(\delta \vec{y}_i) A^+(0)|p \rangle p_i^\alpha p_i^\beta \frac{\partial}{\partial p_i^\alpha} \frac{\partial}{\partial p_i^\beta} = e^{-i\vec{p}_i \cdot \delta\vec{y}_i} \langle p| \partial^\alpha A^+(\delta \vec{y}_i) \partial^\beta A^+(0)|p \rangle \frac{\partial}{\partial p_i^\alpha} \frac{\partial}{\partial p_i^\beta} .
\label{Hexp}
\end{eqnarray}
Since the Taylor expansion imposes the condition $\vec{p}_i \to 0$ on the hard part $H$, it now no longer has functional dependence on $\vec{p}_i$.
Therefore, we may perform the integrations over $\vec{p}_i$ and $\delta\vec{y}_i$.
In this effort, we only keep the leading term arising from the delta functions when evaluating the derivatives on the hard part and ignore the terms from the derivatives on the phase factor $e^{-i {x_{D}}_i p^+ \delta y_i^-}$ which result in spatial moments of the two gluon matrix elements such as $\langle A^+(\delta y^-, \delta \vec{y}_{\perp}) \delta y^- A^+(0) \rangle$.
With the above simplifications, the differential hadronic tensor now reads,
\begin{eqnarray}
\frac{d W_{nn}^{A\mu\nu}}{{d^3l_q} }  \!\!&&\!\! =  W_0^{A\mu\nu}
\frac{1}{n!} \left(\prod_{i=1}^{n} \int_0^{L^-} dY_i^-  \left[ - D_{L1}\frac{\partial}{\partial p_i^-} + \frac{1}{2} D_{L2} \frac{\partial^2}{\partial^2 p_i^-} + \frac{1}{2} D_{T2} {\nabla_{p_{i \perp}}^2}  \right] \right)  \delta^3 \left( \vec{l}_q - \vec{q} - \sum_{i=1}^n \vec{p}_i \right)|_{\vec{p}_1 \cdots \vec{p}_n = 0}. \ \ \ \ \
\end{eqnarray}
{In writing the above expressions, we have focused on the case with unpolarized initial and final states and only kept terms with non-vanishing coefficients by utilizing the symmetry of the system.}
The jet transport coefficients $D_{L1}$, $D_{L2}$ and $D_{T2}$ in the equation above are defined as,
\begin{eqnarray}
D_{L1} \!\!&&\!\! = g^2 \frac{C_F}{N_c^2 - 1} \int dy^- \frac{\rho}{2p^+} \langle p| i\partial^- A^+(y^-) A^+(0)|p \rangle , \nonumber\\
D_{L2} \!\!&&\!\! = g^2 \frac{C_F}{N_c^2 - 1} \int dy^- \frac{\rho}{2p^+} \langle p| \partial^- A^+(y^-) \partial^- A^+(0)|p \rangle,
\nonumber\\
D_{T2} \!\!&&\!\! = g^2 \frac{C_F}{N_c^2 - 1} \int dy^- \frac{\rho}{2p^+} \langle p| \partial_\perp A^+(y^-) \partial_\perp A^+(0)|p \rangle. \ \ \ \ \
\end{eqnarray}
One can immediately see that if the longitudinal momentum exchange is the same order as transverse momentum exchange, then $D_{L2} \sim D_{T2}$ as stated in the Introduction.
This indicates it is important to include both effects when considering the evolution of hard jets in dense nuclear medium.
In what follows, we will demonstrate that these coefficients
$D_{L1}$, $D_{L2}$ and $D_{T2}$ are, up to an overall factor, related to elastic energy loss rate $\hat{e}$,
the diffusion in longitudinal momenta  $\hat{e}_2$ and the diffusion in transverse momenta $\hat{q}$.
{We note that the above definitions of jet transport coefficients are not gauge-invariant. The gauge invariant definition of the transport coefficients can be found with the inclusion of higher order terms where $k_{\perp} \ll \lambda Q$. The summation over an arbitrary number of such
soft gluon insertions leads to the appearance of Wilson lines between the the gluon field strength operators, which renders the operator product
gauge invariant. For the case of jet broadening, more detailed discussion can be found in Ref. \cite{Liang:2008rf, Benzke:2012sz}. }

The resummation over an arbitrary number of multiple scatterings may now be performed,
\begin{eqnarray}
\frac{d W^{A\mu\nu}}{{d^3l_q} } = \sum_{n=0}^{\infty} \frac{d W_{nn}^{A\mu\nu}}{{d^3l_q} }  =  W_0^{A\mu\nu}
\phi(L^-, l_q^-, \vec{l}_{q\perp}),  \ \ \ \ \ \
\end{eqnarray}
where we have defined the final quark distribution function $\phi(L^-, l_q^-, \vec{l}_{q\perp})$,
\begin{eqnarray}
\phi \!\!&&\!\! = \sum_{n=0}^{\infty} \frac{1}{n!} \left(\prod_{i=1}^{n} \int_0^{L^-} dY_i^- \left[ - D_{L1}\frac{\partial}{\partial p_i^-} + \frac{1}{2} D_{L2} \frac{\partial^2}{\partial^2 p_i^-}  + \frac{1}{2} D_{T2} {\nabla_{p_{i \perp}}^2}  \right] \right)   \delta^3 \left( \vec{l}_q - \vec{q} - \sum_{i=1}^n \vec{p}_i \right)|_{\vec{p}_1 \cdots \vec{p}_n = 0}. \ \ \ \ \
\end{eqnarray}
The quark distribution function, after resumming over an arbitrary number of scatterings, reads,
\begin{eqnarray}
\phi \!\!&&\!\! = \exp\left( L^- \left[ D_{L1}\frac{\partial}{\partial l_q^-} + \frac{1}{2} D_{L2} \frac{\partial^2}{\partial^2 l_q^-} + \frac{1}{2} D_{T2} {\nabla_{l_{q\perp}}^2}  \right] \right) \delta(l_q^- - q^-) \delta^2(\vec{l}_{q\perp}),
\end{eqnarray}
where we have converted the derivatives over $p_i$ to the derivatives over $l_q$.

It is easy to see that the above distribution function $\phi(L^-, l_q^-, \vec{l}_{q\perp})$ for the final quark satisfies the following evolution equation,
\begin{eqnarray}
\frac{\partial \phi}{\partial L^-} = \left[ D_{L1}\frac{\partial}{\partial l_q^-} + \frac{1}{2} D_{L2} \frac{\partial^2}{\partial^2 l_q^-} + \frac{1}{2} D_{T2} {\nabla_{l_{q\perp}}^2}  \right] \phi(L^-, l_q^-, \vec{l}_{q\perp}).
\end{eqnarray}
Such an evolution equation describes the evolution of the three-dimensional momentum distribution of propagating partons which suffers multiple scatterings without radiation.
The three terms in the above evolution equation represent the contributions from longitudinal momentum loss and diffusion, and the diffusion of transverse momentum.
Using the initial condition: $\phi(L^- = 0, l_q^-, \vec{l}_{q\perp}) =  \delta(l_q^- - q^-) \delta^2(\vec{l}_{q\perp})$, the distribution $\phi$ has the following solution,
\begin{eqnarray}
\phi = \frac{e^{-{(l_q^- - q^- + D_{L1} L^-)^2}/({2D_{L2}L^-})}}{\sqrt{2\pi D_{L2} L^-}}  \frac{e^{-{{l}_{q\perp}^2 }/({2D_{T2}L^-})}}{2\pi D_{T2} L^-}.
\end{eqnarray}
From the above solution, one may obtain
\begin{eqnarray}
\langle l_q^- \rangle  = q^- - D_{L1} L^-, & \langle (l_q^-)^2 \rangle - \langle l_q^- \rangle^2 = D_{L2} L^-, & \langle l_{q\perp}^2 \rangle  = 2 D_{T2} L^- .
\end{eqnarray}
The coefficients $D_{L1}$, $D_{L2}$ and $D_{T2}$ are defined on the light cone, and they are related to elastic energy loss rate $\hat{e}=dE/dt$, the diffusion of elastic energy loss $\hat{e}_2 = d(\Delta E)^2/dt$, and the transverse momentum diffusion rate $\hat{q} = d(\Delta p_T)^2/dt$ as
\begin{eqnarray}
\hat{e} = D_{L1}, & \hat{e}_2 = D_{L2}/\sqrt{2}, & \hat{q} = 2\sqrt{2} D_{T2}.
\end{eqnarray}

\end{widetext}

\section{Summary}

The current manuscript forms the first of a series of efforts meant to incorporate elastic drag and scattering induced emission into one seamless
formalism.
In this work, we have studied the effect of multiple scattering on the propagation of a single hard parton in the dense nuclear medium.
The developed formalism will be applied to both jet modification in DIS on a large nucleus as well as in the case of heavy-ion collisions.
To study the effect of multiple scattering on a single patron, we have suppressed radiation from the parton. This will be dealt with in a future effort,
which will include the scattering of both the parent parton as well as the radiated gluon.

A class of higher twist corrections which are enhanced by the large extent of the nucleus are resummed to obtain the length dependent
three dimensional momentum distribution of a
hard quark  which exits after traversing the dense medium.
We include both transverse momentum broadening as well as the drag and momentum diffusion in longitudinal direction between the hard jet and medium constituents simultaneously when considering jet propagation in a dense medium. Momentum power counting based on SCET-Glauber scaling indicates that
both these corrections are of similar size, though the longitudinal drag is still very small compared to the energy of the hard quark.
A differential equation has been derived for the time evolution of jet momentum distribution as affected by the multiple scatterings from the dense medium without radiation. The three coefficients involved $D_{L1}, D_{L2}, D_{T2},$ have been related to the elastic drag $\hat{e}$, longitudinal straggling $\hat{e}_{2}$ and
transverse momentum diffusion coefficient $\hat{q}$. The combined effect of all these three coefficients on the radiated gluon spectrum will be explored
in a future effort.

\section*{Acknowledgments}

This work was supported in part by U.S. DOE under grant no DE-FG02-05ER41367 and the NSF under grant no PHY-1207918.


\bibliographystyle{h-physrev5.bst}
\bibliography{GYQ_refs.bib}
\end{document}